\documentclass[reprint,amsmath,amssymb,prl,showpacs]{revtex4-1}

\usepackage{graphicx}
\usepackage{dcolumn}
\usepackage{color}
\usepackage{bm}
\usepackage[colorlinks=true]{hyperref}

\begin{document}

\title{Magnetic Resonance Probing Ensemble  Dynamics}

\author{V.~Herold} \affiliation{Experimentelle Physik V, Julius-Maximilians-Universit{\"a}t W{\"u}rzburg; Germany}
\author{T.~Kampf} \affiliation{Experimentelle Physik V, Julius-Maximilians-Universit{\"a}t W{\"u}rzburg; Germany}
\author{P.M.~Jakob} \affiliation{Experimentelle Physik V, Julius-Maximilians-Universit{\"a}t W{\"u}rzburg; Germany}

\date{\today}

\begin{abstract}
  
   We demonstrate the use of spatially encoded magnetic resonance to quantify ensemble dynamics of microscopic particles below the spatial resolution. By evaluating time series of $k$-space data-points,
  k-dependent motion patterns can be revealed in short measurement time. As no images have to be reconstructed, the proposed method operates directly in the data space of the measurement i.e. the
  $k$-space and allows to examine motion patterns by processing time series of just one $k$-space data-point. To proof the feasibility of this new technique we simulate the MR measurement with samples
  producing particle drift and brownian motion. MR experiments with sedimenting microspheres and rising air-bubbles verify the results of the simulations. This new technique is not limited by
  relaxation times and covers a wide field of applications for particle motion in opaque media.

\end{abstract}

\maketitle

The principles of magnetic resonance build the foundation of a variety of imaging modalities. Generally, images are not acquired directly in the spatial domain but in the reciprocal space, the
$k$-space.  To reconstruct an image, a sufficient number of data points in k-space has to be sampled. Imaging small particle motion in real time (e.g., particles at a size of several micrometers
dispersed in a fluid) can be very challenging if not impossible due to the need of high spatial and temporal resolution. In this paper we show that it is possible to quantify statistical parameters
of particle motion in samples without performing MR-imaging and with very low requirements on the MR-hardware.  This is made possible by transferring the principles known from dynamic light scattering
to MR. This implies that in the simplest case the temporal evolution of only one $k$-space point has to be studied to statistically describe the dynamics of the sample.

During an MR-experiment the magnetization vector of the sample, initially colinear with the main magnetic field ${\bm B_0}=(0,0,B_z)$, is tipped into the transverse plane by applying a resonant
radio-frequency pulse. The transverse magnetization is precessing around ${\bm B_0}$ with the typical Larmor frequency and induces a voltage in a receive coil, which constitutes the measured signal
and decays with a relaxation time of $T_2$. After each RF-pulse the longitudinal component of the magnetization exponentially rebuilds to the thermal equilibrium vector parallel to $\bm B_0$ with a
relaxation time $T_1$. To spatially encode the measurement signal, constant magnetic field gradients are applied $\bm{G}=\nabla B_z$. These gradients are connected to the $k$-value by
$k_i= \gamma\cdot\int G_i d t$, where $\gamma$ is the gyromagnetic ratio. Generally the measurement signal $S(\bm k)$ is sampled in k-space. From a fully sampled k-space signal an image $I(\bm r)$ can be reconstructed by applying the Fourier
transformation:
\begin{equation}
  \label{EQN_-1}
I(\bm{r})=\int S(\bm{k})e^{i \bm{k}\bm{r}} d\bm{k} \, .
\end{equation}

To measure uncorrelated ensemble motion as occurring with diffusion, MR-sequences are additionally expanded with motion-encoding gradient pulses. For instance, in Spin-Echo sequences the
diffusion-sensitizing gradients consist of additional short symmetric gradient pulses (i.e., with amplitude: ${ G}_1={G}_2={ G}$ and pulse duration: $\delta_1=\delta_2=\delta$) arranged before and
after the $180^{\circ}$-pulse \cite{stejskal1965}. Using the narrow gradient pulse approximation, Callaghan et al. first revealed the formal analogy between the so called Pulsed Gradient Spin Echo
(PGSE)-MR and neutron scattering \cite{Callaghan1984a}. If one defines a reciprocal space vector ${\bm q}=\gamma \delta {\bm G}$ it can be shown, that the incoherent part of the expectation value for
the signal of distributed magnetic moments can be written as:

 \begin{equation}
  \label{EQN_0}
S(\bm{q},\Delta)=N^{-1}  \left\langle \sum_{i=1}^N e^{-i\bm{q}[\bm{r}_i(\Delta)-\bm{r}_i(0)]} \right\rangle \, ,
\end{equation}
where $\Delta$ represents the temporal spacing between the diffusion gradient pulses and N is the number of the precessing magnetic moments \cite{callaghan2011translational}. PGSE-MR was also applied
to image diffusion in granular flow \cite{Seymour2000}. Laun et al. could recently show, that by breaking the symmetry between the two diffusion-encoding gradients the diffusion experiment can be
shifted from being a scattering experiment towards being an imaging experiment and thus enabling to define the shape of the diffusion boundaries which are in general not accessible by a pure
scattering experiment \cite{Laun2011,Seymour2000}. Classical resolution limits in MR-imaging, as defined by a limited magnitude of the maximum $k$-space vector, can thus be circumvented.

In this work we propose a method to study the motion of particles below the spatial resolution of conventional imaging, by evaluating time series of single data points acquired in the reciprocal space
($k$-space). This approach generalizes the idea of interpreting dynamic data acquisition in $k$-space as a scattering experiment without the necessity of image reconstruction in analogy to concepts
used in dynamic light scattering (DLS) and differential dynamic microscopy \cite{berne1976,pedersen2002,Cerbino2008}. Let us assume a sample made of a number of identical particles dispersed in a
viscous fluid. It is irrelevant whether the source for the signal is found in the particles or in the fluid, provided that the particles generate sufficient signal contrast to the fluid background to
create variations in the spatial signal distribution. For convenience, in the following discussion the particles are supposed to generate the signal instead of the surrounding fluid (which in
generally would be the case for MR). The spatial distribution of the signal density $F(\bm{r}, t)$ can then be written as the convolution \cite{pedersen2002}.:
 
\begin{equation}
  \label{EQN_1}
  F(\bm{r}, t)=F_0(\bm{r})\ast \sum_{j} \delta [\bm{r}-\bm{r} _j(t)] \, .
\end{equation}

$F_0$ represents the local spatial distribution of the particle signal analogously to the scattering potential of a single particle as known from optics.  The time-dependent $k$-space signal for the
moving particles, that is generated by applying constant magnetic field gradients can be derived from \eqref{EQN_1} by applying the spatial Fourier-transformation \eqref{EQN_-1}:

\begin{equation}
  \label{EQN_2}
  f(\bm{k}, t)=f_0(\bm{k})\cdot \sum_{j} e^{i \bm{k}\bm{r}_j(t)} \, .
\end{equation}

The autocorrelation function of the $k$-space signal corresponds to the field correlation function in dynamic light scattering:

\begin{eqnarray}
 \Gamma({\bm k},\Delta t) &=& \langle f({\bm k},t) \cdot f^{\ast}({\bm k},t+\Delta t)  \rangle_t \nonumber \\
                                         & =&  |f_0({\bm k}) |^2 \left \langle \sum_{m,n} e^{i {\bm k}    [    {\bm r}_m (t+\Delta t)-    {\bm r}_n (t)   ]  }
                                              \right \rangle_t \,  .
  \label{EQN_4}
\end{eqnarray}

In DLS experiments it is accessible through the {\it Siegert} relation \cite{pedersen2002}. In MR-measurements the field correlation function can be directly calculated from the time course of a
single $k$-space point.  For uncorrelated and identical particles its normalized version can be written as \cite{pedersen2002}:
\begin{equation}
  g(\bm{k},  \Delta  t)= \left \langle e^{i {\bm k}    [    {\bm r}(t+\Delta t)-    {\bm r} (t)   ]  }  \right \rangle_t \, .
\label{EQN_5}
\end{equation}

When assuming identical particles Eq.\ \eqref{EQN_5} directly corresponds to the signal attenuation in a PGSE-experiment given in Eq.\ \eqref{EQN_0}. Instead of dephasing the measured signal by
applying diffusion-encoding gradients and subsequently exploiting its effect on the signal attenuation one can also simply repeat single-point $k$-space sampling and evaluate the temporal correlation
in $k$-space. The concept of extracting dynamic information by acquiring time-series of $k$-space data is already applied in dynamic MR-imaging, primarily to reconstruct undersampled data
\cite{Tsao2003}. The novelty of the proposed method however is the description of $k-t$ data in a scattering framework operating on just one $k$-space position and thus forgoing the reconstruction of
MR-images. The temporal distance between the diffusion gradients $\Delta$ in Eq.\ \eqref{EQN_0} corresponds to $\Delta t$ in the field correlation function (Eq.\ \eqref{EQN_5}). Albeit $\Delta$ must not exceed
$T_2$ (or $T_1$ in stimulated echo experiments), $\Delta t$ in Eq.\ \eqref{EQN_5} can virtually be extended to infinity since there is no theoretical upper limit for the repetition time ($TR$) in an
MR-experiment. This enables to also quantify very slow statistical motion.

From dynamic light scattering the course of the field correlation function Eq.\ \eqref{EQN_5} is well known for Brownian motion as an exponential decay \cite{pedersen2002}

\begin{equation}
  g(\bm{k}, \Delta t)= e^{-D\bm k^2 \Delta t} \, ,
\label{EQN_6}
\end{equation}
with the diffusion coefficient $D$ and for a constant particle drift as an oscillating function
\begin{equation}
  g(\bm{k}, \Delta t)= e^{i {\bm v}_0 \bm k\Delta t}  \, ,
\label{EQN_8}
\end{equation}
with the drift velocity ${\bm v}_0$. In cases where most of the signal energy is contained in the static background signal, it is more favorable to examine functions describing the statistics of the
signal differences in order to cancel out the background signal \cite{Giavazzi2009}. One of these functions is the averaged mean square of the signal differences in $k$-space also known as the
structure function:
\begin{equation}
  S'(\bm{k}, \Delta t)= N^{-1}\left \langle |f({\bm k},t+\Delta t) - f({\bm k},t) |^2  \right \rangle_t \, ,
\label{EQN_10}
\end{equation}
where N is the number of particles. The time course of the  structure function is closely related to the field correlation function. The normalized version of the  structure function  can be written as:
\begin{equation}
  S(\bm{k}, \Delta t)= \frac{ S'({\bm k}, \Delta t)}{ S'({\bm k},+\infty)} =1- \Re[g({\bm k}, \Delta t) ] \, ,
\label{EQN_11}
\end{equation}
where $\Re[f]$ denotes the real part of $f$. The structure function allows to extract the sample dynamics from the fluctuating $k$-space signal even in cases of strong background signal.

\begin{figure}
\includegraphics[scale=0.95]{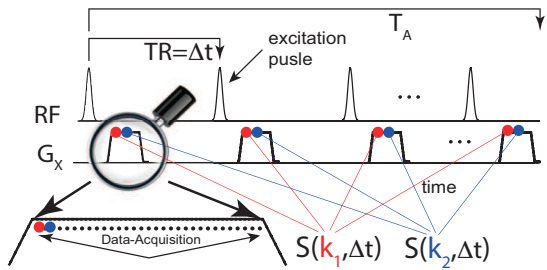}
\caption{\label{fig:sequence} Repetitive applied excitation-pulse and gradient-pulse module to generate a time series of spatially encoded signal  in k-space. }
\end{figure}

\begin{figure*}
\includegraphics[scale=0.95]{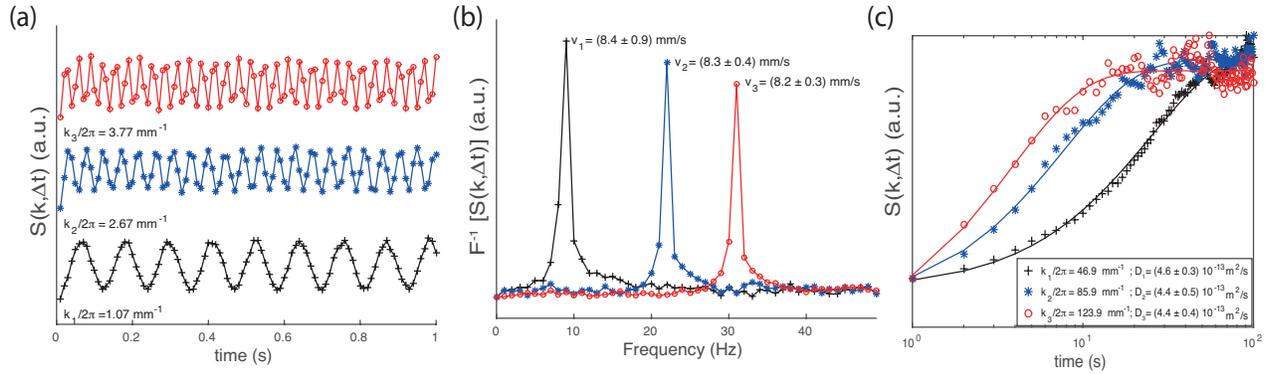}
\caption{\label{fig:simulation}Simulation: a) Structure function for constant drifting microspheres ($\O \, 100 \, \mu m$) at different $k_x$-values. The oscillating time course reflectes the constant drift
velocity of the sample. b) Corresponding spectra, with calculated velocity values for each frequency peak. c) Structure function for Brownian motion of $1 \mu \, m$ particles in water (at
300 K). The results of the simulation correspond well with the theoretically calculated diffusion coefficient  $D\,=\, 4.39\cdot 10^{-13} \, \mathrm{m^2/s}$. }
\end{figure*}

To prove the feasibility of the presented method two different MR experiments are simulated. First, the signal for a sample based on glass spheres dispersed in water ($\rho= \mathrm {2500 \, kg/m^3}$,
$\O \,= 100 \,\mu m $) is simulated with a constant gravitational drift motion according Stokes Law set to $v$=8.175 mm/s. The simulated MR-sequence only consists of a 1D frequency-encoded readout (
$G_R=0.08 \, \mathrm{T/m}$ ) along the particle drift direction as shown in Fig.\ \ref{fig:sequence}. Full spoiling is simulated by clearing the transverse part of the magnetization vector after each
sequence repetition. Data acquisition is performed as shown in Fig.\ \ref{fig:sequence}. For each time-series of $k$-space-data points the structure function is evaluated according to Eq.\
\eqref{EQN_10} and Eq.\ \eqref{EQN_11}. Although sampling just one $k$-space data point under the readout gradient would be sufficient, 128 sampling points are acquired (each represents a different
$k$-space position) with a sampling frequency of 100 kHz. With this data it can be validated, if the structure functions deliver comparable results for different $k$-space values. The sequence is
repeated 100 times at a repetition time of 10 ms such that the measurement of one experiment covers a time window of 1s. Fig.\ \ref{fig:simulation}a shows the time course of the normalized structure
function for three different $k$-space values. The oscillating progression of the normalized structure function clearly reflects the translational movement of the particles at a constant
velocity. Identifying the peaks in the discrete spectra as shown in Fig. \ref{fig:simulation}b, allows to determine the drift velocity. The velocities determined from the peak positions ($v_1$=(8.4
$\pm$ 0.9) mm/s; $v_2$=(8.3 $\pm$ 0.4) mm/s; $v_3$=(8.2 $\pm$ 0.3) mm/s) show a good agreement with the predefined drift velocity of 8.2 mm/s of the particles.
In a second step we simulate Brownian motion of 1 $\mu$m particles dispersed in water (viscosity $\eta=1.0 \cdot 10^{-3} \,\mathrm {Ns/m^2} $; temperature T=300 K) using a 3D-random walk with a time
resolution of $\tau=\mathrm 10^{-5} \mathrm{s}$ and a stepsize of $\langle x \rangle =\sqrt{2D \tau}$ for each direction. According to the Stokes-Einstein-relation the diffusion coefficient is
calculated to be \mbox{$D\,=\, 4.39\cdot 10^{-13} \,\mathrm{m^2/s}$}. To adapt the MR-sequence to the slow diffusion process the following changes are made to the sequence: $G_R=0.23 \, \mathrm{T/m}$,
repetition time $TR= 1 \, \mathrm{s}$, sampling frequency 10 kHz, number of sequence repetitions 100. \mbox{Fig.\ \ref{fig:simulation}c} shows the structure function plotted for three different
$k$-values and fitted to the exponential decay as given in Eqs.\ \eqref{EQN_6} and \eqref{EQN_11}. The results of the simulation show a good agreement with the given diffusion coefficient
$D\,=\, 4.39\cdot 10^{-13} \mathrm{m^2/s}$ ($D_1\,=\, (4.6 \,\pm\, 0.3) 10^{-13} \, \mathrm{m^2/s}$, $D_2\,=\, (4.4 \,\pm\, 0.5)10^{-13} \,\mathrm{m^2/s}$,
$D_3\,=\, (4.4 \,\pm\, 0.4)10^{-13} \ \mathrm{m^2/s}$). Both simulations are designed to resample {\it real world}-experiments, that is gradient strengths and sequence-timings can easily be realized
with existing MR-hardware.

As a proof of principle, in a first MR-experiment, we examine sedimenting microspheres ($\rho= \mathrm {2500 \, kg/m^3}$, $\O \,= 100 \,\mu m $) on a Buker \mbox{17.6 T} scanner equipped with a
\mbox{1 T/m} gradient insert. The temporal evolution of the $k$-space-data is acquired using a 1D fast gradient-echo sequence i.e. without phase-encoding ($k_y$=0) \cite{haase1986}. The sequence is
repeated $\mathrm{N_t}=200$ times with a repetition time of $TR=3.6 \,\mathrm{ms}$ to capture the sample dynamics in a total time window of $T_A=0.7$s. The direction of the frequency-encoding
($k_x$-encoding) is chosen to be aligned with the drift direction. Under each readout-gradient 64 $k_x$-sampling-points were acquired ($-0.7 \, \mathrm{mm}^{-1} <k_x/2 \pi < 2.0 \,
\mathrm{mm}^{-1}$). For averaging purposes, the whole experiment was repeated $\mathrm{N_{av}}=64$ times.

Additionally, images are acquired with a 2D fast gradient-echo sequence with the same frequency encoding gradients as above (now also using phase-encoding); imaging parameters are as follows: flip-angle
$\alpha = 10^\circ$, slice-thickness 1 mm; field-of -view (xy): $(25 \time 25 \times 25) \, \mathrm{mm}^2$ and resolution (xy): $0.4 \times 0.4 \, \mathrm{mm}^2$.

\begin{figure*}
\includegraphics[scale=0.9]{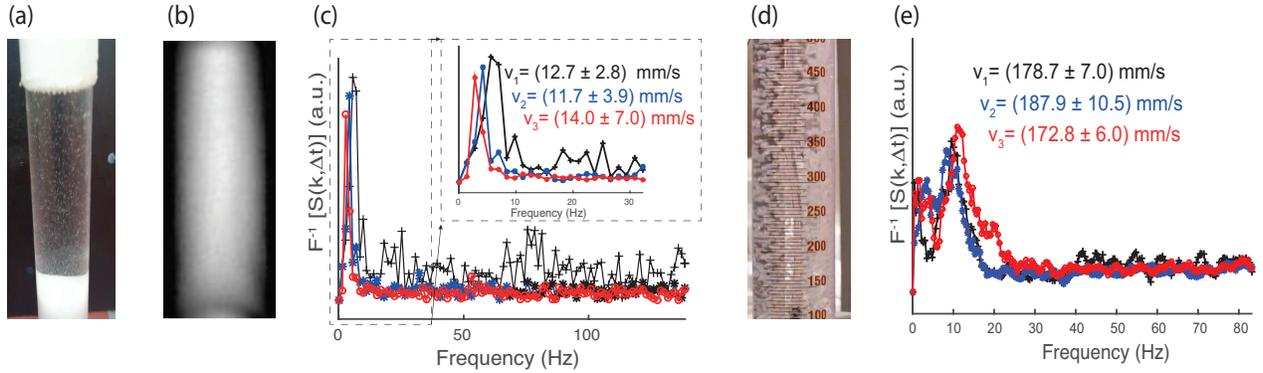}
\caption{\label{fig:MR_EXP} Experiments: a) Snapshot of the drifting particels. b) Reconstructed MR-image: single particles can not be resolved. c) Frequency spectrum of the structure functions
  showing the velocity distribution of sedimenting glass spheres. The given values represent the calculated drift velocity at each frequency peak. Errors represent the limited spectral resolution. d)
  Snapshot of rising air bubbles in a glass tube filled with water. e) Frequency spectrum of the moving bubbles.}
\end{figure*}

Fig.\ \ref{fig:MR_EXP}a shows a photograph of the sedimenting glass spheres. Due to the limited spatial resolution and primarily due to the limited temporal resolution a reconstructed MR-image reveals
no details about the particle size and distribution as shown in Fig.\ \ref{fig:MR_EXP}b.  When picking out the time course of one $k_x$-encoded data point, the imprint of the ensemble motion becomes
visible, which can be seen in the corresponding spectra of the structure functions as shown in Fig.\ \ref{fig:MR_EXP}c. The evaluation of the spectra for three different $k_x$-values yields a mean
velocity value of (12.8 $\pm$ 1.2) mm/s. When analyzing high-resolution video frames of the sedimenting glass spheres, a mean drift-velocity of (11.4 $\pm$ 0.9) mm/s could be determined, which is in
good agreement with the MR-experiments.

In a second experiment the velocity distribution of rising air-bubbles in a water tube is examined. The experiments are conducted with a Magnetom Skyra-3T (Siemens). $k$-space data is acquired again
using a 1D fast gradient-echo sequence, i.e., without phase-encoding (repetition time $TR= 6 \,\mathrm{ms}$; $\mathrm{N_{t}}=383$). Under each readout-gradient 80 $k_x$-sampling-points are acquired
($-0.14 \, \mathrm{mm}^{-1} <k_x/2 \pi < 0.14 \, \mathrm{m}^{-1}$). The whole experiment is averaged $\mathrm{N_{av}}=32$ times.  As air-bubble source a conventional bubble air pump is used, supplied
with a bubble diffuser as commonly found in aquariums. With video analysis and single air-bubble tracking we could identify a range of velocities between 150 and 250 mm/s. In Fig.\ \ref{fig:MR_EXP}e
the corresponding spectra of the structure functions are plotted for three different $k_x$-values. For each $k_x$-value the velocity at the most prominent peak is given. The width of the spectral
distribution shows a good agreement with the range of velocities that were found by video analysis.

Both examples show the capability of the presented method to examine particle dynamics.  The modulation of the field correlation function reflects the dynamics of the particles and is governed by the
magnitude of the corresponding $k$-vector. Even if measuring very low diffusion coefficients or very small drift velocities sufficient signal dynamics can
be generated by expanding the temporal sampling window without exceeding the limits of the maximum gradient fields. This is possible since the single data acquisition steps can be set to any temporal
distance not limited by the relaxation time.  The proposed method is very robust to modifications of the $k$-space signal, e.g., caused by field inhomogeneities, since the quantitative evaluation of the
signal amplitude is not required to examine the sample dynamics. It can be applied to virtually any MR system with at least a 1D gradient system and  seems to be applicable in all fields
where the dynamics of ensemble motion in opaque media are of interest, such as statistics of cell dynamics, long term drift or diffusion experiments.  

We gratefully acknowledge Volker Behr and Patrick Vogel for helpful discussions and careful reading of the manuscript.


\begin{thebibliography}{8}%
\makeatletter
\providecommand \@ifxundefined [1]{%
 \@ifx{#1\undefined}
}%
\providecommand \@ifnum [1]{%
 \ifnum #1\expandafter \@firstoftwo
 \else \expandafter \@secondoftwo
 \fi
}%
\providecommand \@ifx [1]{%
 \ifx #1\expandafter \@firstoftwo
 \else \expandafter \@secondoftwo
 \fi
}%
\providecommand \natexlab [1]{#1}%
\providecommand \enquote  [1]{``#1''}%
\providecommand \bibnamefont  [1]{#1}%
\providecommand \bibfnamefont [1]{#1}%
\providecommand \citenamefont [1]{#1}%
\providecommand \href@noop [0]{\@secondoftwo}%
\providecommand \href [0]{\begingroup \@sanitize@url \@href}%
\providecommand \@href[1]{\@@startlink{#1}\@@href}%
\providecommand \@@href[1]{\endgroup#1\@@endlink}%
\providecommand \@sanitize@url [0]{\catcode `\\12\catcode `\$12\catcode
  `\&12\catcode `\#12\catcode `\^12\catcode `\_12\catcode `\%12\relax}%
\providecommand \@@startlink[1]{}%
\providecommand \@@endlink[0]{}%
\providecommand \url  [0]{\begingroup\@sanitize@url \@url }%
\providecommand \@url [1]{\endgroup\@href {#1}{\urlprefix }}%
\providecommand \urlprefix  [0]{URL }%
\providecommand \Eprint [0]{\href }%
\providecommand \doibase [0]{http://dx.doi.org/}%
\providecommand \selectlanguage [0]{\@gobble}%
\providecommand \bibinfo  [0]{\@secondoftwo}%
\providecommand \bibfield  [0]{\@secondoftwo}%
\providecommand \translation [1]{[#1]}%
\providecommand \BibitemOpen [0]{}%
\providecommand \bibitemStop [0]{}%
\providecommand \bibitemNoStop [0]{.\EOS\space}%
\providecommand \EOS [0]{\spacefactor3000\relax}%
\providecommand \BibitemShut  [1]{\csname bibitem#1\endcsname}%
\let\auto@bib@innerbib\@empty



\bibitem [{\citenamefont {Stejskal}\ and\ \citenamefont {Tanner}(1965)}]{stejskal1965}%
  \BibitemOpen \bibfield {author} {\bibinfo {author} {\bibfnamefont {E.}~\bibnamefont {Stejskal}}\ and\ \bibinfo {author} {\bibfnamefont {J.}~\bibnamefont {Tanner}},\ }\href@noop {} {\bibfield
    {journal} {\bibinfo {journal} {J Chem Phys}\ }\textbf {\bibinfo {volume} {42}},\ \bibinfo {pages} {288} (\bibinfo {year} {1965})}\BibitemShut {NoStop}%
\bibitem [{\citenamefont {Callaghan}(1984)}]{Callaghan1984a}%
  \BibitemOpen \bibfield {author} {\bibinfo {author} {\bibfnamefont {P.}~\bibnamefont {Callaghan}},\ }\href@noop {} {\bibfield {journal} {\bibinfo {journal} {Australian Journal of Physics}\ }\textbf
    {\bibinfo {volume} {37}},\ \bibinfo {pages} {359} (\bibinfo {year} {1984})}\BibitemShut {NoStop}%
\bibitem [{\citenamefont {Callaghan}(2011)}]{callaghan2011translational}%
  \BibitemOpen \bibfield {author} {\bibinfo {author} {\bibfnamefont {P.~T.}\ \bibnamefont {Callaghan}},\ }\href@noop {} {\emph {\bibinfo {title} {Translational dynamics and magnetic resonance:
        principles of pulsed gradient spin echo NMR}}}\ (\bibinfo {publisher} {Oxford University Press},\ \bibinfo {year} {2011})\BibitemShut {NoStop}%
\bibitem [{\citenamefont {Seymour}\ \emph {et~al.}(2000) \citenamefont {Seymour}, \citenamefont {Caprihan}, \citenamefont {Altobelli},\ and\ \citenamefont {Eiichi}}] {Seymour2000}%
  \BibitemOpen \bibfield {author}{\bibinfo {author} {\bibfnamefont {J.-D.}~\bibnamefont {Seymour}}, \bibinfo {author} {\bibfnamefont {A.}~\bibnamefont {Caprihan}}, \bibinfo {author} {\bibfnamefont
    {S.-A.}~\bibnamefont {Altobelli}}, \ and\ \bibinfo {author} {\bibfnamefont {E.}\ \bibnamefont {Fukushima}},\ }\href@noop {} {\bibfield {journal} {\bibinfo {journal} {Phys Rev Lett}\ }\textbf
  {\bibinfo {volume} {84}},\ \bibinfo {pages} {266} (\bibinfo {year} {2000})}\BibitemShut {NoStop}%
\bibitem [{\citenamefont {Laun}\ \emph {et~al.}(2011)\citenamefont {Laun}, \citenamefont {Kuder}, \citenamefont {Semmler},\ and\ \citenamefont {Stieltjes}}]{Laun2011}%
  \BibitemOpen \bibfield {author} {\bibinfo {author} {\bibfnamefont {F.~B.}\ \bibnamefont {Laun}}, \bibinfo {author} {\bibfnamefont {T.~A.}\ \bibnamefont {Kuder}}, \bibinfo {author} {\bibfnamefont
      {W.}~\bibnamefont {Semmler}}, \ and\ \bibinfo {author} {\bibfnamefont {B.}~\bibnamefont {Stieltjes}},\ }\href@noop {} {\bibfield {journal} {\bibinfo {journal} {Physical review letters}\ }\textbf
    {\bibinfo {volume} {107}},\ \bibinfo {pages} {048102} (\bibinfo {year} {2011})}\BibitemShut {NoStop}%
\bibitem [{\citenamefont {Berne}\ and\ \citenamefont {Pecora}(1976)}]{berne1976}%
  \BibitemOpen \bibfield {author} {\bibinfo {author} {\bibfnamefont {B.~J.}\ \bibnamefont {Berne}}\ and\ \bibinfo {author} {\bibfnamefont {R.}~\bibnamefont {Pecora}},\ }\href@noop {} {\emph {\bibinfo
      {title} {Dynamic light scattering: with applications to chemistry, biology, and physics}}}\ (\bibinfo {publisher} {Courier Corporation},\ \bibinfo {year} {1976})\BibitemShut {NoStop}%
\bibitem [{\citenamefont {Cerbino}\ and\ \citenamefont {Trappe}(2008)}]{Cerbino2008}%
  \BibitemOpen \bibfield {author} {\bibinfo {author} {\bibfnamefont {R.}~\bibnamefont {Cerbino}}\ and\ \bibinfo {author} {\bibfnamefont {V.}~\bibnamefont {Trappe}},\ }\href@noop {} {\bibfield
    {journal} {\bibinfo {journal} {Phys Rev Lett}\ }\textbf {\bibinfo {volume} {100}},\ \bibinfo {pages} {188102} (\bibinfo {year} {2008})}\BibitemShut {NoStop}%
\bibitem [{\citenamefont {Pedersen}(2002)}]{pedersen2002}%
  \BibitemOpen \bibfield {author} {\bibinfo {author} {\bibfnamefont {J.}~\bibnamefont {Pedersen}},\ }\href@noop {} {\enquote {\bibinfo {title} {Neutrons, x-rays and light. scattering methods applied
        to soft condensed matter, edited by p.  lindner \& t. zemb},}\ } (\bibinfo {year} {2002})\BibitemShut {NoStop}%
\bibitem [{\citenamefont {Tsao}\ \emph {et~al.}(2003)\citenamefont {Tsao}, \citenamefont {Boesiger},\ and\ \citenamefont {Pruessmann}}]{Tsao2003}%
  \BibitemOpen \bibfield {author} {\bibinfo {author} {\bibfnamefont {J.}~\bibnamefont {Tsao}}, \bibinfo {author} {\bibfnamefont {P.}~\bibnamefont {Boesiger}}, \ and\ \bibinfo {author} {\bibfnamefont
      {K.~P.}\ \bibnamefont {Pruessmann}},\ }\href {http://dx.doi.org/10.1002/mrm.10611} {\bibfield {journal} {\bibinfo {journal} {Magn Reson Med}\ }\textbf {\bibinfo {volume} {50}},\ \bibinfo {pages}
    {1031} (\bibinfo {year} {2003})}\BibitemShut {NoStop}%
\bibitem [{\citenamefont {Giavazzi}\ \emph {et~al.}(2009)\citenamefont {Giavazzi}, \citenamefont {Brogioli}, \citenamefont {Trappe}, \citenamefont {Bellini},\ and\ \citenamefont
    {Cerbino}}]{Giavazzi2009}%
  \BibitemOpen \bibfield {author} {\bibinfo {author} {\bibfnamefont {F.}~\bibnamefont {Giavazzi}}, \bibinfo {author} {\bibfnamefont {D.}~\bibnamefont {Brogioli}}, \bibinfo {author} {\bibfnamefont
      {V.}~\bibnamefont {Trappe}}, \bibinfo {author} {\bibfnamefont {T.}~\bibnamefont {Bellini}}, \ and\ \bibinfo {author} {\bibfnamefont {R.}~\bibnamefont {Cerbino}},\ }\href
  {http://pre.aps.org/abstract/PRE/v80/i3/e031403} {\bibfield {journal} {\bibinfo {journal} {Physical Review E}\ }\textbf {\bibinfo {volume} {80}},\ \bibinfo {pages} {031403} (\bibinfo {year}
    {2009})}\BibitemShut {NoStop}%
\bibitem [{\citenamefont {Haase}\ \emph {et~al.}(1986)\citenamefont {Haase}, \citenamefont {Frahm}, \citenamefont {Matthaei}, \citenamefont {Hanicke},\ and\ \citenamefont {Merboldt}}]{haase1986}%
  \BibitemOpen \bibfield {author} {\bibinfo {author} {\bibfnamefont {A.}~\bibnamefont {Haase}}, \bibinfo {author} {\bibfnamefont {J.}~\bibnamefont {Frahm}}, \bibinfo {author} {\bibfnamefont
      {D.}~\bibnamefont {Matthaei}}, \bibinfo {author} {\bibfnamefont {W.}~\bibnamefont {Hanicke}}, \ and\ \bibinfo {author} {\bibfnamefont {K.-D.}\ \bibnamefont {Merboldt}},\ }\href@noop {}
  {\bibfield {journal} {\bibinfo {journal} {Journal of Magnetic Resonance (1969)}\ }\textbf {\bibinfo {volume} {67}},\ \bibinfo {pages} {258} (\bibinfo {year} {1986})}\BibitemShut {NoStop}%
\end{thebibliography}
\end{document}